\documentclass[12pt]{article}
\usepackage{graphicx}

\newcommand\pubnumber{SNSN-323-63}
\newcommand\pubdate{\today}

\def\support{\footnote{This work was supported by the U. S. Department of Energy Office of Science, Office of Nuclear Physics under Award Number DE-FG02-97ER-41014,.}}

\def\Title#1{\begin{center} {\Large #1 } \end{center}}
\def\Author#1{\begin{center}{ \sc #1} \end{center}}
\def\Address#1{\begin{center}{ \it #1} \end{center}}

\newcommand\pubblock{\rightline{\begin{tabular}{l} \pubnumber\\
         \pubdate  \end{tabular}}}
\newenvironment{Abstract}{\begin{quotation}  }{\end{quotation}}
\newenvironment{Presented}{\begin{quotation} \begin{center} 
             PRESENTED AT\end{center}\bigskip 
      \begin{center}\begin{large}}{\end{large}\end{center} \end{quotation}}
\def\Acknowledgements{\bigskip  \bigskip \begin{center} \begin{large}
             \bf ACKNOWLEDGEMENTS \end{large}\end{center}}

\def\beq{\begin{equation}}
\def\eeq#1{\label{#1}\end{equation}}
\def\eeqn{\end{equation}}

\def\beqa{\begin{eqnarray}}
\def\eeqa#1{\label{#1}\end{eqnarray}}
\def\eeqan{\end{eqnarray}}

\let\bar=\overbar

\def\Dslash{\not{\hbox{\kern-4pt $D$}}}
\def\dslash{\not{\hbox{\kern-2pt $\del$}}}

\def\msb{{\bar{\ssstyle M \kern -1pt S}}}

\begin{document}
\begin{titlepage}
\pubblock

\vfill
\Title{The Proton Radius Puzzle- Why We All Should Care}
\vfill
\Author{ Gerald A. Miller\support}
\Address{Physics Department, University of Washington, Seattle, Washington 98195-1560, USA }
\vfill
\begin{Abstract}
The  status of the proton radius puzzle (as of the date of the Conference) is reviewed. The most likely potential theoretical and experimental explanations are discussed. Either the electronic hydrogen experiments were not sufficiently accurate to measure the proton radius, the two-photon exchange effect was not properly accounted for, or there is some kind of new physics. I expect that  upcoming experiments  will resolve this issue within the next year or so.

\end{Abstract}
\vfill
\begin{Presented}
Conference on the Intersections between particle and nuclear physics,   \\
Indian Wells, USA,  May 29-- June 3, 2018
\end{Presented}
\vfill
\end{titlepage}
\def\thefootnote{\fnsymbol{footnote}}
\setcounter{footnote}{0}

This title is chosen because understanding of the proton radius puzzle requires knowledge of atomic, nuclear and particle physics. 
The puzzle began with the publication of the results of the 2010 muon-hydrogen experiment in 2010~\cite{pohl} and its confirmation~\cite{Antognini:1900ns}. The proton radius ($r_p^2=-1/6  G_E'(Q^2=0)$  was measured to be 
$ r_p =0.84184 (67)$   fm, which contrasted with the value obtained from electron spectroscopy  $  r_p =0.8768 (69)$ fm. This   difference of about 4\% has become known as the proton radius puzzle~\cite{Miller:2011yw}.  We  use the technical terms: the radius 0.87 fm is denoted as large, and the one of 0.84 fm as small.
At that time the large  value was consistent with that obtained (with much larger uncertainties) from electron scattering. The significant  recent discussion of extraction of $r_p$ from electron scattering will not be addressed here, except to  remark that very different values of $r_p$ have been obtained from the same data set. The PRAD experiment~\cite{Gasparian:2014rna} plans to remove this problem by making measurements in the range $7.7\times 10^{-3}<Q^2<0.13 $ fm$^{-2}$, which may be sufficient for an accurate extraction.

One might wonder whether or not a 4\% difference really matters. After all, 4\% is pretty small and the value of $r_p$ cannot be calculated to that accuracy. The real issue is whether or not the fundamental electron-proton interaction is the same as the muon-proton interaction. There have been hints that violations of the  principle of lepton-universality exist, and the issues discussed here may be part of a much wider picture.

This subject has been reviewed in 2013~\cite{Pohl:2013yb} and 2015~\cite{Carlson:2015jba}. The present talk discusses more recent experiments, and the attempts to explain the proton radius puzzle.
\begin{figure}[htb]
\centering
\includegraphics[height=2.5in]{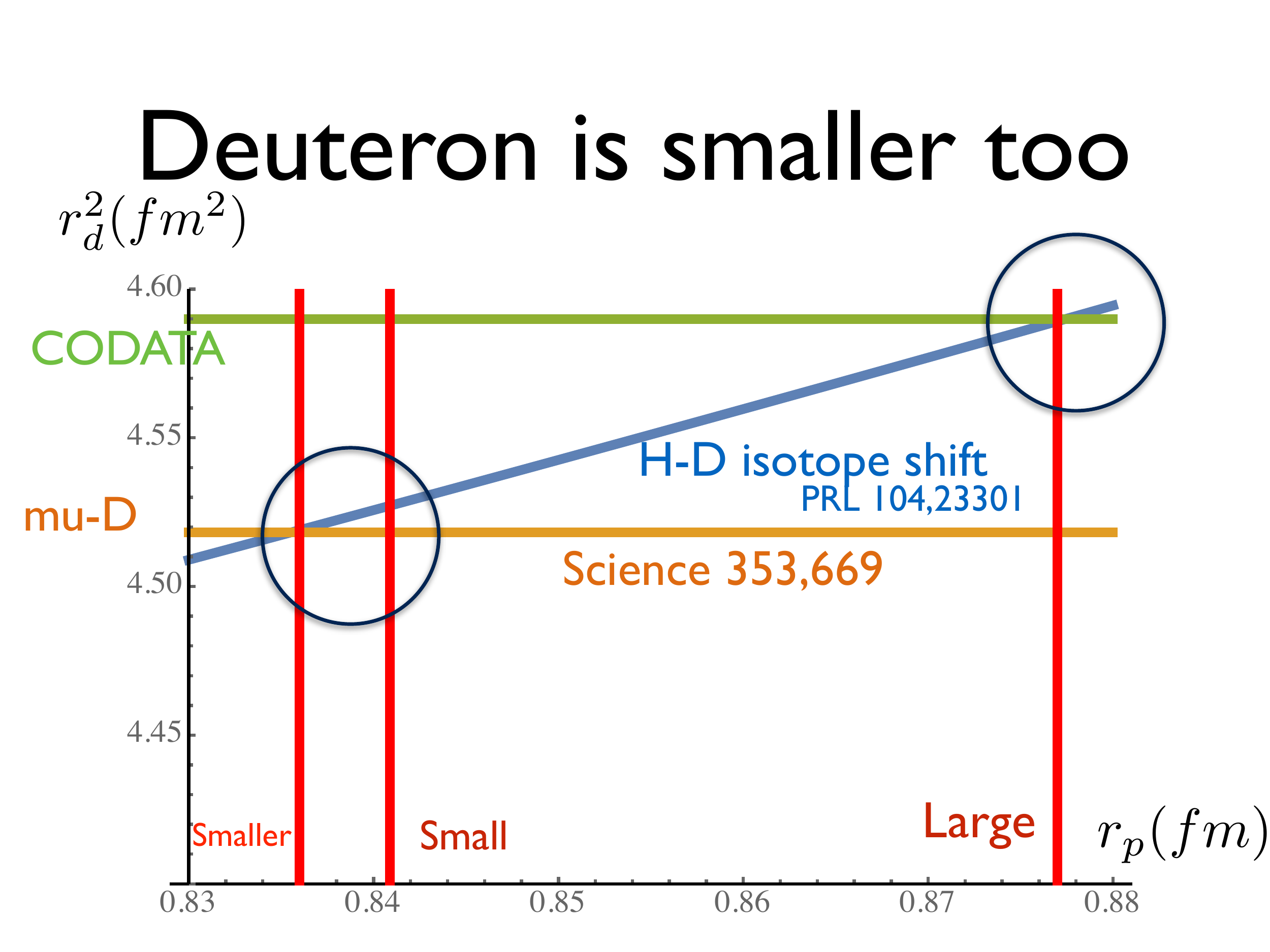}
\caption{Deuteron radius {\it vs.} proton radius The green horizontal line shows the value of $r_d^2$ as analyzed by the CODATA collaboration. The orange horizontal line shows the muonic-deuteron value, which is smaller. The Hydrogen-Deuteron isotope shift experiment~\cite{Parthey:2010aya} (blue line) provides the relation between $r_d^2$ and $r_p$. The vertical red lines denote the large and small values of the proton radius. The fact that the red, blue and orange lines do not intersect at the same point shows  a discrepancy in the result that has become known as the deuteron radius puzzle.}
\label{fig:deut}
\end{figure}
\section{Experiments- H, D, $^3$He, $^4$He} In a conference talk in  2014 (Mainz, 2 June 2014) R. Pohl reported that 
the proton radius $r_p=0.8409\pm0.0004$ fm as measured with muons and $r_p=0.8770\pm 0.045$ fm as measured by an average of  electron inputs. That difference is 7.9 standard deviations.

It is also known that the deuteron radius as measured by muons~\cite{Pohl1:2016xoo}
is smaller also. This is seen  Fig.~\ref{fig:deut}. 

Preliminary data for muonic $^3$He and $^4$He have been reported~\cite{Antognini:2015moa}. The measured radii are consistent with expected values. However, as of the date of CIPANP18, the uncertainties for $^3$He were relatively large. Those for $^4$He are very small and present a serious challenge for any theory for the proton radius puzzle.

Thus the summary so far is:  
\begin{itemize}
\item there is a very significant  difference between the muonic  and electronic  hydrogen spectroscopy  determinations of the   the proton charge radius

\item there is a  similar difference  in the deuteron, not totally arising from the difference in proton radii, implying an  effect  for the  neutron charge radius
\item there is no radius puzzle for  $^3$He (large error bars)
\item there is no effect in $^4$He (small error bars)
\item any explanation of the proton radius puzzle must account for the above, assuming that  there is a difference between muon and electron hydrogen spectroscopy
\end{itemize}
\subsection{New results from electronic hydrogen}
Beyer et al \cite{beyer:2017} determined the proton charge radius from the $2s-4p$ transition finding the small radius in agreement with the results of muonic spectroscopy. However, the Paris group~\cite{Fleurbaey:2018fih} measured the $1s-3s$ transition in electronic hydrogen and found the large radius. This difference is unsettling as it raises the question of whether or not electron-spectroscopy is sufficient to determine the proton charge radius  to the desired accuracy. Two transitions are needed to determine both the Rydberg constant and the Lamb shift. Both of these experiments also use the $1s-2s$ energy difference which tightly controls the value of  the Rydberg constant.  However, another experiment is in progress (E.~A.~Hessels, conference talk): an original Lamb shift measurement of the $2s-2p$ energy difference that doesn't rely on another transition.This might be decisive.  

Possible resolutions of the proton radius puzzle include:
(a) Electronic  H muonic spectroscopic  determinations of the proton charge radius really do not disagree. 
(b)  There is some missing strong interaction effect in the two photon exchange diagram that is responsible for the different determinations~\cite{Miller:2012ne}.
(c) The  muon interacts differently than electron! {\it eg.}~\cite{TuckerSmith:2010ra,Izaguirre:2014cza}.

The remainder of this document shifts from the history to our efforts to explain the proton radius puzzle.
\section{How much is needed?}
Suppose one wants to account for the apparent difference between electronic and muonic determinations of the proton radius without requiring that the radii are actually different. This accounting can be done by finding a previously unaccounted effect in the energy difference.  Going back the the original determination~\cite{pohl}, the measured energy difference, 206.2942(32) meV, is written as a sum of three terms:
\begin{eqnarray}
206.2942(32)=206.0573(45)-5.2262 r_p^2+0.0347 r_p^3 \,{\rm meV}.
\end{eqnarray}
One can account for the difference in radii by increasing the computed value of the radius-independent contribution, 206.0573 meV, by a 0.31 meV attractive effect on the 2S state. 
So one can try to resolve the proton radius puzzle  finding a new effect of that size and sign. The ideas in the literature include two-photon exchange effects  and new particles.  
\section{Two-photon Exchange}
\begin{figure}[htb]
\centering
\includegraphics[height=2.5in]{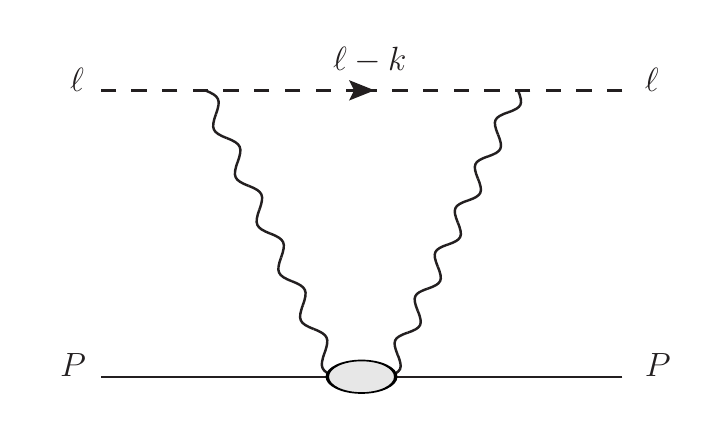}
\caption{Two photon exchange}
\label{2g}
\end{figure}

Two-photon exchange is an interesting effect because it produces an effect proportional to the fourth power of the lepton mass. Thus it can be significant for muonic hydrogen, but insignificant for electronic hydrogen. Fig.~\ref{2g} shows the relevant Feynman diagram. The energy shift depends on the real part of the virtual forward Compton scattering amplitude.   The imaginary part is measured, so one uses dispersion relations to get the needed real part.  There are two spin-independent amplitudes known as $T_{1,2}$. But $T_1(\nu,Q^2)$ does not fall fast enough as $\nu$ increases towards infinity to allow Cauchy's theorem to be used without using a subtraction. Thus one needs to use a function $T_1(0,Q^2)$ that is not measurable for $Q^2\ne0$. 

Previous work~\cite{Miller:2012ne} found that there is enough freedom in this function to account for the proton radius puzzle and not be in conflict with the computed proton-neutron mass difference, which  also depends on $T_1(0,Q^2)$.  
The computed energy shift reproduces the necessary 0.31 meV, and also produced a measurable effect in muon-proton scattering. 

However, the two-photon exchange effect also leads to a big contribution to the Lamb shift for $^4$He.
This is because the computed  Lamb shift for a nucleus of $Z$ protons and $N$ neutrons that arises from any hadronic effect is proportional to  $ Z^3(Z \delta E^p_L+N \delta E^n_L)$~\cite{Miller:2015yga}
where $\delta E^{p,n}_L$ is a new contribution to the Lamb shift for a proton (p) or neutron (n).  Given $E^p_L=0.31 $ meV, one finds a shift of 4.8 meV for $^4$He, which is surely ruled out by the measured Lamb shift.

Nevertheless, the two-photon exchange effect was~\cite{pohl}  the largest uncertainty in the precision muonic  Lamb shift proton measurement, and it remains the largest uncertainty.

The two-photon exchange effect will be studied in the MUSE (muon scattering experiment)~\cite{Gilman:2013eiv} .  This experiment will measure $e^\pm$ and $\mu^\pm$ scattering on the proton.
 
This experiment constrains the  two photon effect, which still survives at a  significant level. 
If the large radius is correct, and  the two photon effect is accounted for, all leptons will the see the large radius.
If the small  radius is correct, and  the  two photon effect is accounted for, all leptons will the see the small radius.
This experiment cannot detect the presence of a new light scalar particle, but  if such exists and the two photon effect is accounted for, all leptons will  see the large radius. 
\section{Electrophobic scalar boson}
 There is another muonic anomaly~\cite{Gorringe:2015cma}. This is in the well-known difference between the measured and computed muonic magnetic moment. One can postulate a new scalar boson that accounts for the missing magnetic moment. It is noteworthy that the inclusion of a new scalar boson can account for the missing Lamb shift of 0.31 meV and the missing magnetic moment.

The postulated scalar boson would be exchanged between a lepton and  proton.
The requirements for such a particle include: (a)
reproduce  the $\mu-p$  Lamb shift
(b) cause almost no hyperfine  effect in the muon-proton atom
(c)
produce a small effect for the deuteron, and  almost no effect in $^4$He
(d) be consistent with the anomalous magnetic moments of the muon and electron
(e)  avoid many other constraints
(f)  be found.
Items (a)-(e) have been covered.

These issues are discussed in the papers~\cite{Liu:2015sba,Liu:2016mqv,Liu:2017htz,Liu:2018qgl}.  We explain briefly how this works~\cite{Liu:2015sba}.
A new scalar boson,    
 $\phi$, that couples to the muon and proton could explain both the proton radius and $(g-2)_\mu$ puzzles~\cite{TuckerSmith:2010ra}. We investigated the couplings of this boson to standard model fermions, $f$, which appear as terms in the Lagrangian, $\mathcal{L}\supset e \epsilon_f\phi\bar ff$, where $\epsilon_f=g_f/e$ and $e$ is the electric charge of the proton.  Other authors   pursued this    idea, but  
  made further assumptions relating the couplings to different species; e.g. in~\cite{TuckerSmith:2010ra}, $\epsilon_p$ is taken equal to $\epsilon_\mu$ and in \cite{Izaguirre:2014cza}, mass-weighted couplings are assumed. References~\cite{TuckerSmith:2010ra} and~\cite{Izaguirre:2014cza} both neglect $\epsilon_n$.       We made no {\it a priori} assumptions regarding signs or magnitudes of the coupling constants. The Lamb shift in muonic hydrogen fixes $\epsilon_\mu$ and $\epsilon_p$ to have the same sign. $\epsilon_e$ and $\epsilon_n$ are allowed to have either sign.

We focus on the scalar boson possibility because scalar exchange produces no hyperfine interaction, in accord with observation~\cite{pohl,Antognini:1900ns}.
 The emission of possible 
new vector particles becomes copious at high energies, and in the absence of an ultraviolet completion, is ruled out~\cite{Karshenboim:2014tka}.

Scalar boson exchange can account for  both  the proton radius puzzle and the $(g-2)_\mu$ discrepancy~\cite{TuckerSmith:2010ra}. The shift of the lepton $(\ell=\mu,e)$ muon's magnetic moment due to one-loop $\phi$ exchange is  {given by}~\cite{Jackiw:1972jz}.
 Scalar exchange between fermions $f_1$ and $f_2$ leads to a Yukawa potential, 
 \begin{equation} V(r)=-\epsilon_{f_1}\epsilon_{f_2}\alpha e^{-m_\phi r}/r.\end{equation} We used
$\Delta a_\mu=287(80)\times 10^{-11},\, \delta E_L^{\mu\rm H}=-0.307(56)~\rm meV
$
within two standard deviations. This value of  $\delta E_L^{\mu\rm H}$,  is the same as the energy shift caused by using the different values of the proton 
radius~\cite{pohl,Antognini:1900ns}  to explain the two discrepancies. This allowed us to determine   both $\epsilon_p$ and $\epsilon_\mu$ as functions of $m_\phi$. 

We  used the  constraints: (a)  low energy scattering of neutrons  on $^{208}$Pb~\cite{Leeb:1992qf}, sensitive to the couplings of the scalar to neutrons, $\epsilon_n$, and protons, $\epsilon_p$,   to  obtain new 
bounds on $m_\phi$. Allowing independent values of $\epsilon_n$ and $\epsilon_p$ increased  the allowed parameter space. (b) The known NN charge-independence breaking scattering length difference. (c) The binding energy of nuclear matter (d) The difference in the binding energies of $^3$He and $^3$H  and (e)  the preliminary results on the Lamb shifts in muonic deuterium and muonic $^4$He.

Using these observables we  limited the ratio of the coupling of $\phi$ to neutrons and protons, $\epsilon_n/\epsilon_p$. If the couplings to neutron and proton   are of opposite sign, they   interfere destructively, masking the effects of the  $\phi$ and substantially weakening the limits on the magnitudes of $\epsilon_n,\epsilon_p$.
For a given value of $\epsilon_n/\epsilon_p$, we 
 used the shift of the binding energy in $N=Z$ nuclear matter and the difference in binding energies of $^3$H and $^3$He to constrain $\epsilon_p$.   

We  explored the coupling of the scalar to electrons, which is of particular experimental importance because electrons are readily produced. We found limits on the coupling $\epsilon_e$  that are very strict. 
 
In addition to muonic atoms, scalar exchange will affect the Lamb shift in ordinary electronic atoms. To set limits on the coupling, 
 we required that the change to the Lamb shift in hydrogen is $\delta E_L^{\rm H}<14~\rm kHz$~\cite{Eides:2000xc}(2 S.D.). 
The publications~\cite{Liu:2015sba,Liu:2016mqv,Liu:2017htz,Liu:2018qgl} show the  allowed regions of parameter space.

The latest work along these lines used the additional constraints provided by $\eta$ decay~\cite{Liu:2018qgl} to find that  the electron beam dump experiments limit the  allowed range of $m_\phi$ to  between about 200 keV and 3 MeV. This narrow range represents an inviting target for ruling out or discovering this scalar boson. 
 
\section{Summary}
This talk was concerned with the main resolutions of the proton radius puzzle. Either the electronic hydrogen experiments were not sufficiently accurate to measure the proton radius, the two-photon exchange effect was not properly accounted for, or there is some kind of new physics. I expect that  upcoming experiments  will resolve this issue within the next year or so.

\Acknowledgements
 
I would like to thank the MIT LNS, the Southgate Fellowship of Adelaide University, the Batheba de Rothchild Fellowship of Hebrew University, the Shaaoul Fellowship of Tel Aviv University, the Physics Division of Argonne National Laboratory and the U.S. Department of Energy Office of Science under award number DE-FG02-97ER-41014 for support that enabled this work.

\end{document}